# A Robust Image Watermarking System Based on Deep Neural Networks

Xin Zhong*
University of Nebraska at Omaha
xzhong@unomaha.edu

Frank Y. Shih
New Jersey Institute of Technology
shih@njit.edu

*Abstract*—Digital image watermarking is the process of embedding and extracting watermark covertly on a carrier image. Incorporating deep learning networks with image watermarking has attracted increasing attention during recent years. However, existing deep learning-based watermarking systems cannot achieve robustness, blindness, and automated embedding and extraction simultaneously. In this paper, a fully automated image watermarking system based on deep neural networks is proposed to generalize the image watermarking processes. An unsupervised deep learning structure and a novel loss computation are proposed to achieve high capacity and high robustness without any prior knowledge of possible attacks. Furthermore, a challenging application of watermark extraction from camera-captured images is provided to validate the practicality as well as the robustness of the proposed system. Experimental results show the superiority performance of the proposed system as comparing against several currently available techniques.

*Index Terms*—Image watermarking, robustness, deep learning, convolutional neural networks, phone camera scan

## I. Introduction

DIGITAL image watermarking refers to the process of embedding and extracting information covertly on a carrier image. The data (i.e., the watermark) is hidden into a cover-image to create a marked-image that will be distributed over the Internet. However, only the authorized recipients can extract the watermark information correctly. According to user's demands, the watermark can be in different forms, for instances, some random bits or electronic signatures for image protection and authentication as well as some hidden messages for covert communication [1]. The watermark can be encoded for different purposes, such as increasing the perceivable randomness for additional security via encryption methods or restoring the impact of noise via error correction codes for watermark integrity under attacks [2, 3].

While the primary concentration of a steganographic system is the imperceptibility to human vision as well as the undetectability to computer analysis, an image watermarking system often controls the robustness as its priority. Thus the watermark should survive even if the marked-image is degraded or distorted [4]. Ideally, a robust image watermarking system keeps the watermark intact under a designated class of distortions without the assistance of other techniques. However, in practice the robust image watermarking systems often extract the watermark approximately under malicious attacks and apply various encoding methods for restoration [5, 6].

Traditional image watermarking schemes manually design algorithms for the watermark embedding and extraction. For example, the least significant bits (LSB) based strategies place the watermark on a cover-image through bits substitutions or other mathematical operations [5, 7]. Although the trivial replacement enables the invisibility, LSB-based methods are less robust and can be easily revealed by statistical analysis. More advanced watermarking schemes place the watermark on various image domains. For example, Cox *et al*. [8] embedded the watermark on the frequency spectrum for high fidelity and high security. Shih and Zhong [9] increased the frequency domain capacity while preserving the fidelity. Pevny *et al*. [10] enhanced the security by an embedding scheme that maintains the cover image statistics. Zong *et al*. [11] improved the robustness by embedding the watermark into image histogram.

Incorporating deep neural networks with image watermarking has attracted increasing attention during recent years. In contrast to significant achievements in steganalysis for hidden data reveal [12, 13], very few attempts applying deep learning in watermark embedding and extraction are reported. Earlier methods [14-16] used neural networks to assign the significance for the bits of each pixel instead of manual determination. Tang *et al*. [17] proposed a generative adversarial network to determine the embedding position and the strength on the cover-image. Kandi *et al*. [18] used two deep autoencoders for non-blind binary watermark extraction in the marked-image, where the pixels produced by the first auto-encoder represent bit zero and the pixels produced by the second auto-encoder represent bit one. Baluja *et al*. [19] applied deep autoencoders for blind image watermarking to achieve high fidelity as well as high capacity. Li *et al*. [20] embedded the watermark in the discrete cosine domain and used convolutional neural networks for extraction. However, due to fragility of deep neural networks [21], the robustness issue becomes a challenge since inputting a modified image to a pre-trained deep learning system can cause failure. Mun *et al.* [22] proposed adversarial networks to solve this issue by including attack simulation in the training.

Developing robust image watermarking systems for watermark extraction from camera resamples requires that the watermark must simultaneously resist multiple distortions, such as geometric distortions, optical tilt, quality degradation, compression, lens distortions, and lighting variations [23, 24]. Researchers have developed various methods in solving these problems. Katayama *et al.* [25] proposed a sinusoidal watermark pattern for robust watermark embedding and a visible frame for marked-image rectification. Other methods based on the autofocus function of a phone camera have been developed, such as embedding the watermark through a correlation function, placing the watermark in selected positions via spread spectrum, and applying log-polar transformation [26-28]. Pramila *et al.* [24] proposed watermark extraction from a camera resample of an image printed on blank paper by combining computational photography and robust image watermarking, but the nonblind property of the system restricts its application range.

In this paper, we develop an automated image watermarking system using deep learning networks based on three main motivations. First, exploring the fitting ability of deep learning models in learning the rules of watermark embedding is helpful in developing an automated system. Second, the proposed system is tested on the application of watermark extraction from camera resamples, providing a potential solution towards this challenging issue. Third, image watermarking is viewed from a novel perspective – an image fusion task [29, 30] between the cover-image and the latent spaces of the watermark, where the fused result (i.e., the marked-image) contains the watermark while references the visual appearance of the cover-image.

The remainder of this paper is organized as follows. The proposed system is presented in Section 2. Experiments and analyses are described in Section 3. The application of watermark extraction using a phone camera to scan a screen is given in Section 4. Finally, conclusions are drawn in Section 5.

## II. THE PROPOSED SYSTEM

### A. Preliminaries

Fig. 1 shows a general image watermarking system. The watermark $w$ is inserted into the cover-image $c$ to generate a marked-image $m$ that will be transported through a communication channel. The receiver extracts the watermark data $w^*$ from the received marked-image $m^*$, which may be a modified version of $m$ if some distortions or attacks are occurred during transmission. A robust image watermarking system intends to secure the integrity of the watermark, i.e., minimizing the difference between $w$ and $w^*$.

Conventional strategies formulate an image watermarking task as preserving certain parts from the cover-image for the watermark. As given in Eq. (1), $w$ is embedded by taking some proportions in a domain of $c$,

$$m = \alpha D(c) + \beta w \qquad (1)$$

where $\alpha$ and $\beta$ are the weights which control the watermark strength and $D(c)$ denotes an image domain of the cover-image. Different optimization schemes can be applied to control the embedding and enable the extraction of $w^*$ from $m^*$ according to user's purposes. Some keys, as in cryptographic systems, can also be used in generating, embedding, or extracting the watermark for various applications and extra protections [5].

In contrast, we view image watermarking as an image fusion task. Given two input spaces of the watermark and the cover-image, $W = R^{D_1}$ and $C = R^{D_2}$. The input watermark space is firstly mapped to one of its latent spaces (a feature space $W_f = R^{d_1}$) by a function $\mu: W \to W_f$, and then the watermark embedding is performed by a mapping function $\sigma: \{W_f, C\} \to M$ that fuses the feature space of the watermark and the input cover-image space to produce an intermediate latent space $M = R^{d_2}$. $M$ is the space of the marked-image with two main constraints. The visual appearance of $M$ must be similar to $C$, while the feature of $M$ must correlate to the feature of $W_f$. Therefore, $M$ has the desired attributes of marked-images. On the other hand, watermark extraction is performed by two mapping functions, $\varphi: M \to W_f$ that reconstructs the feature space $W_f$ from $M$, and $\gamma: W_f \to W$ that reconstructs the watermark data from $W_f$.

### B. Overall Architecture

We apply deep neural networks $\mu_{\theta_1}$, $\sigma_{\theta_2}$, $\varphi_{\theta_3}$ and $\gamma_{\theta_4}$ with parameters $\theta_1$, $\theta_2$, $\theta_3$ and $\theta_4$ to learn the mapping functions $\mu$, $\sigma$, $\varphi$ and $\gamma$. The architecture of the proposed image watermarking system is shown in Fig. 2, where $w_i$, $w_f^i$, $c_i$, and $m_i$ are the examples of the spaces $W$, $W_f$, $C$ and $M$. $\mu_{\theta_1}$ and $\sigma_{\theta_2}$ are named as the embedder network, and $\varphi_{\theta_3}$ and $\gamma_{\theta_4}$ are named as the extractor network.

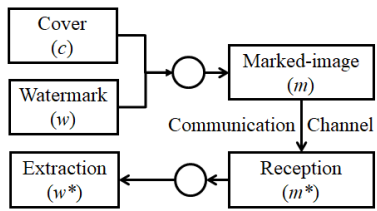

Fig. 1. A general image watermarking system.

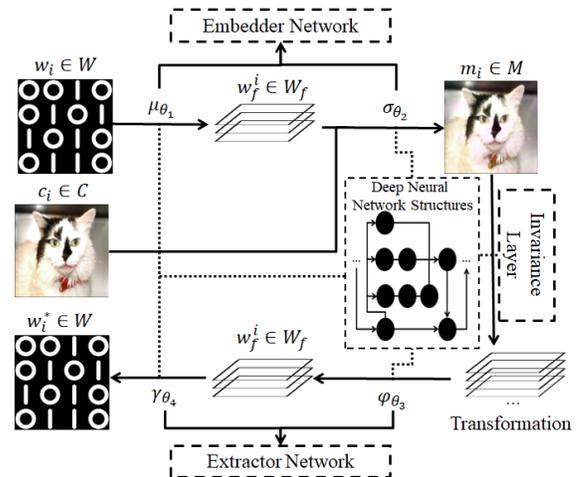

Fig. 2. The architecture of the proposed system.

By taking two inputs, the embedder network transforms the input spaces $W$ and $C$ to the intermediate space $M$. Instead of assigning some unnoticeable portions of the visual components as the watermark, $\sigma_{\theta_2}$ learns to replace the visual appearance of $W_f$ with $C$ while maintaining the characteristics of $W_f$. Hence, the space $M$ after the fusion contains the information from $W$ and $C$. On the contrary, the extractor network takes in a transformation of $M$ and learns to separate and reconstruct $W_f$ and $W$. The overall structure of the proposed system is compatible with the unsupervised deep autoencoders [31], in which an input space can be transformed to a latent space containing the most representative features. The original input can be recovered from the latent space. Similarly, the proposed system transforms two input spaces to a desired latent space and reconstruct one of the inputs from the latent space. The recovery ability of the autoencoders, that ensures an exact reconstruction of the input with appropriate features extracted by the deep neural networks, can secure the feasibility of the proposed structure. The blindness property is enabled since the reconstruction only takes from the latent space, and the fidelity is enabled by the constraints placed on the learned latent space. A latent space in autoencoders is often learned through a bottleneck for the dimensionality compression, while the proposed system learns over-complete representations for both accurate watermark reconstruction and robustness.

The entire system is trained as a single deep neural network. In this presentation, the samples of space $C$ are considered as $128 \times 128 \times 3$ color images. The watermark is assumed to be binary data that could be raw or encoded of 1024-bit information (reshaped to $32 \times 32$). Hence, the presented system has a fixed capacity of 1kb.

*C. Invariance Layer*

To tolerate the distortions on the marked-images without considering all possible attacks, an invariance layer is developed to reject irrelevant information. The invariance layer introduces a function $\tau: M \to T$ that maps space $M$ to an over-complete transformation space $T$. The neurons in this layer are activated in a sparse manner not only to allow a possible loss in $M$ for robustness but also to enhance computational efficiency. As shown in Fig. 3, it converts a 3-channel instance $m_i$ of $M$ into an $N$-channels ($N \geq 3$) instance $t_i$ of $T$ by a fully-connected layer, where $N$ is the redundant parameter. Increasing $N$ means higher redundancy in $T$, which implies higher tolerance of the errors in $M$ and thus enhancing the robustness.

Referring to the contractive autoencoder [32], the invariance layer employs a regularization term to achieve the sparse activation that is obtained by the Frobenius norm of the Jacobian matrix of the layer outputs with regards to the training inputs. Mathematically, the regularization term $P$ is given as

$$P = \sum_{i,j}\left(\frac{\partial h_j(X)}{\partial X_i}\right)^2 \quad (2)$$

where $X_i$ denotes the $i$-th input and $h_j$ denotes the output of the $j$-th hidden unit. Similar to the common gradient computation in neural networks, the Jacobian matrix can be written as

$$\frac{\partial h_j(X)}{\partial X_i} = \frac{\partial A(\omega_{ji}X_i)}{\partial \omega_{ji}X_i}\omega_{ji} \quad (3)$$

where $A$ is an activation function and $\omega_{ji}$ is the weight between $h_j$ and $X_i$. The hyperbolic tangent (tanh) is applied as the activation function of the invariance layer for strong gradients as well as bias avoidance [33]. With $A$ being assigned as the hyperbolic tangent, $P$ can be computed as

$$P = \sum_j(1 - h_j^2)^2 \sum_i(\omega_{ji}^T)^2 \quad (4)$$

Minimizing term $P$ alone essentially renders the weights in the layer unchangeable to all the inputs $X$. However, placing it as a regularization in the total loss computation enables the layer to preserve only useful information while rejecting all other noises and irrelevant information to achieve the robustness.

Different from the contractive autoencoder, each channel in $m_i$ is treated as a single input in the invariance layer to improve the computational efficiency. For example, treating one pixel in $m_i$ as an input means 49,152 inputs for a $128 \times 128 \times 3$ marked-image. Setting the redundant parameter $N$ as its smallest value 3 will imply 147,456 units in the fully-connected invariance layer, which requires at least 7,247,757,312 parameters. This is not practical in most of the current graphic computation units and significantly lowers the efficiency. On the contrary, treating one channel as an input unit considers only 3 input units for the RGB marked-image, which enables faster computation as well as a much larger $N$ for higher robustness.

*D. Embedder and Extractor Network Structure*

Taking the samples $w_i$ from the space $W$, $\mu_{\theta_1}$ with the parameter $\theta_1$ learns a mapping from $W$ to its feature space $W_f$, and $\gamma_{\theta_4}$ learns the reverse mapping of $W_f$ to $W$ with samples $w_f^i$. As shown in Fig. 4, the structures of $\mu_{\theta_1}$ and $\gamma_{\theta_4}$ are symmetric. In $\mu_{\theta_1}$, the $32 \times 32 \times 1$ binary watermark samples are successively increased into $32 \times 32 \times 24$ and $32 \times 32 \times 48$ by each of two convolution blocks. The result reshaped to $128 \times 128 \times 3$ is the feature space sample $w_f^i$. Reversely, $\gamma_{\theta_4}$ reshapes the $128 \times 128 \times 3$ $w_f^i$ back to $32 \times 32 \times 48$, and successively decreases it to a $32 \times 32 \times 1$ binary watermark.

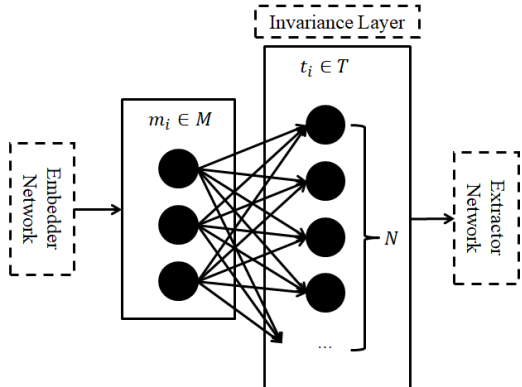

Fig. 3. The invariance layer.

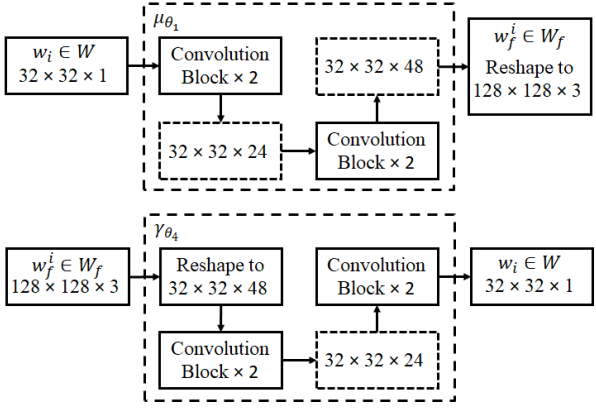

Fig. 4. $\mu_{\theta_1}$ and $\gamma_{\theta_4}$.

Obviously, the space $W$ is increased by 48 times and then restored. The purpose of the increment can be summarized into two-fold. First, it produces a $w_f^i$ that has the same size as the cover-image sample $c_i$ to facilitate a concatenation step in $\sigma_{\theta_2}$. Second, the increment in the latent space $w_f^i$ introduces some redundancy, decomposition, and perceivable randomness to $w_i$, which not only helps robustness but also provides additional security. A few $32 \times 32$ binary watermark samples and their corresponding $128 \times 128 \times 3$ samples from $W_f$ are shown in Fig. 5.

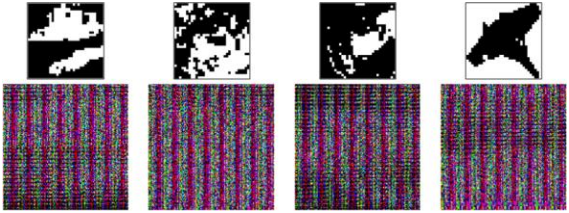

Fig. 5. Samples of the space $W$ and $W_f$. First row: samples from $W$, and second row: their corresponding samples from $W_f$.

To partition the patterns in the binary watermark into different channels, the inception residual block [34] is adopted as the convolution block in the proposed system. It consists of a $1 \times 1$, a $3 \times 3$, a $5 \times 5$ convolution, and a residual connection that sums up the features and the input itself, so that various perception fields are included in the feature extraction. In the proposed structure, each convolution has 32 filters, and the $5 \times 5$ convolution is replaced by two $3 \times 3$ convolutions for efficiency. These 32-channel features are concatenated along the channel dimension to form a 96-channel feature, and a $1 \times 1$ convolution is applied to convert the 96-channel feature back to the original input channels for the summation in the residual connection. Fig. 6 presents a convolution block $f$, where $F_1$, $F_2$, and $F_3$ denote the height, width, and the channel of the block input, respectively.

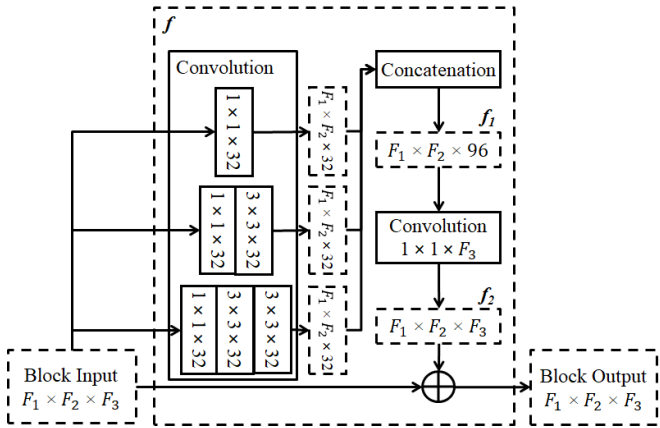

Fig. 6. A convolution block $f$.

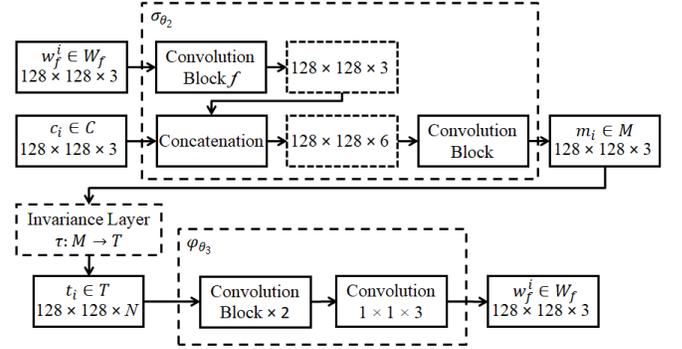

Fig. 7. $\sigma_{\theta_2}$ and $\varphi_{\theta_3}$.

By taking the samples $w_f^i$ from the space $W_f$ along with the samples $c_i$ from the space $C$, the $\sigma_{\theta_2}$ with the parameter $\theta_2$ learns to fuse these two spaces to obtain the marked-image space $M$. Reversely, $\varphi_{\theta_3}$ learns to detect and extract $W_f$ from the transformation space $T$ of $M$. As shown in Fig. 7, the convolution block $f$ is firstly used to extract $w_f^i$ features that are concatenated along the channel dimension with the cover-image sample $c_i$. Another convolution block takes the $128 \times 128 \times 6$ concatenation and fuses it to generate the space $M$. To achieve the fidelity, $M$ contains the feature of $W_f$ while referencing the visual contents of $C$. On the other hand, $\varphi_{\theta_3}$ takes in the $128 \times 128 \times N$ transformation sample $t_i$ produced by the invariance layer and maps it back to $w_f^i$ by two convolution blocks.

Instead of using the space $W_f$, the proposed structure fuses the feature space of $W_f$ obtained through the convolution block $f$ into the space $C$ with the main purpose of controlling the appearance of the space $M$. Visually, the intermediate latent space $M$ should primarily rely on the components of $C$, so the input sample $c_i$ is directly exploited in the structure. In contrast, the information of $w_f^i$ should not be displayed on $m_i$, and hence the feature of $m_i$ is designed to be correlated to the feature of $w_f^i$. This indirect fusion enables the fidelity in the proposed system. In summary, space $M$ borrows the visual contents from $C$ and preserves the features from $W_f$. Various samples of $C$, $M$ and $W_f$ are shown in Fig. 8. Human vision can hardly tell the differences between marked- and cover-images in the spatial domain, while the convolution blocks in $\varphi_{\theta_3}$ are able to find and extract $w_f^i$.

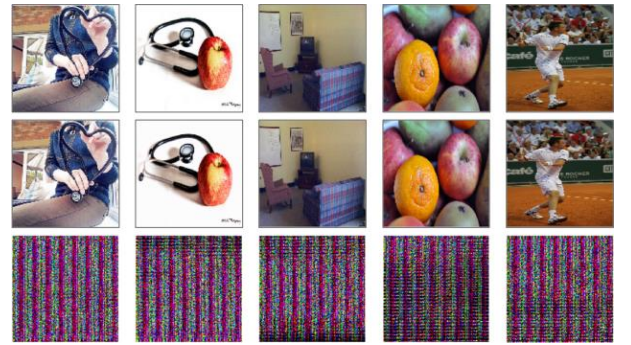

Fig. 8. Samples of the space $C$, $M$ and $W_f$. First row: samples from $C$, second row: samples from $M$, and third row: the corresponding original and extracted samples from $W_f$.

*E. System Objective*

The proposed system intends to learn the mapping functions $\mu$, $\sigma$, $\varphi$, $\gamma$ and $\tau$, using the neural networks $\mu_{\theta_1}$, $\sigma_{\theta_2}$, $\varphi_{\theta_3}$, $\gamma_{\theta_4}$ and $\tau_{\theta_5}$ parametrized by $\theta_1$, $\theta_2$, $\theta_3$, $\theta_4$ and $\theta_5$ given the data samples including $w_i \in W$ and $c_i \in C$. The proposed system is trained as a single deep neural network with a few constraints. Like the autoencoders, the system maps the space $W$ to itself. Hence, the ground truth of $W$ is $W$ itself, and the distance between the input $w_i$ and the system output $w_i^*$ must be minimized. What dissimilar to the autoencoders is that the intermediate latent space $M$ in the proposed system is an image that looks similar to the input space $C$, but contains features extracted from $W$. For this purpose, the system minimizes the distance between the generated samples of the intermediate latent space $m_i$ and the samples of the input space $c_i$, while maximizes the correlation between the samples from the feature space of $W_f$ and the samples from the feature space of $m_i$. Denoting the parameters to be learned as $\vartheta = [\theta_1, \theta_2, \theta_3, \theta_4, \theta_5]$, the empirical risk $L(\vartheta)$ of the proposed system can be expressed as

$$L(\vartheta) = \frac{1}{B}\sum_{i=1}^{B}[||w_i^* - w_i||_1 + ||m_i - c_i||_1 + \psi(m_i, w_f^i)] \quad (5)$$

where $B$ is the number of training examples and $\psi$ is a function computing the correlation as given below.

$$\psi(m_i, w_f^i) = \frac{1}{2}(||g(f_1(w_f^i)), g(f_1(m_i))||_1 \\ + ||g(f_2(w_f^i)), g(f_2(m_i))||_1)$$
(6)

where $g$ denotes the Gram matrix of all possible inner products. Besides $w_f^i$, the convolutional block $f$ in $\sigma_{\theta_2}$ also extracts features from $m_i$, and the correlation between these features is maximized by minimizing the distance between the Gram matrices. To highlight the overall performance rather than a few outliers, the mean absolute error is selected to compute the distance.

Along with the regularization $P$ computed by Eq. (4), the structural risk of the proposed model can be represented as $L(\vartheta) + \lambda P$, where $\lambda$ is the weight controlling the strength of the regularization term. The objective of the system is to learn the parameter $\vartheta^*$ that minimizes the structural risk.

$$\vartheta^* = \mathrm{argmin}_\vartheta L(\vartheta) + \lambda P \quad (7)$$

In the gradient flow during the backpropagation in the training, the term $||w_i^* - w_i||_1$ is applied by all the components of the proposed structure in their weights updates, while only the embedder network ($\mu_{\theta_1}$ and $\sigma_{\theta_2}$) applies term $||m_i - c_i||_1$ and $\psi(m_i, w_f^i)$ to their weight updates.

## III. EXPERIMENTS AND ANALYSES

*A. Training and Testing*

By providing a fixed watermarking capacity of 1,024 bits, the proposed system is trained using ImageNet [35] (rescaled to $128 \times 128$) as the cover-image and the binary version of CIFAR [36] ($32 \times 32$) as the watermark. Both datasets include more than millions of images to introduce a large scope of instances to the system. The ADAM [37] optimizer that applies a moving window in gradient computation is adopted for its ability of continuous learning after large epochs. Fig. 9 shows the value of the terms in the empirical risk and of the structural risk during 200 epochs. At the training and testing, both T1 and T2 in $L(\vartheta)$ converge smoothly below 1.5% and $L(\vartheta) + \lambda P$ converges below 3%. Term T1 has slightly more errors because there are some modifications on the marked-image to indicate the watermark features. $\lambda$ is set to be 0.01 in this case, and all the layers in the system apply the rectified linear unit (ReLU) as the activation function except that $m_i$ and $w_i^*$ use sigmoid to limit the output range into (0, 1) and the invariance layer uses hyperbolic tangent.

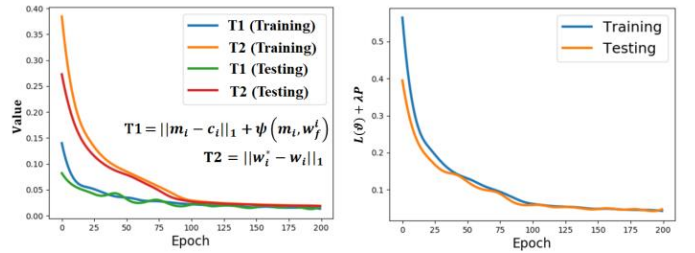

Fig. 9. The empirical risk and the structural risk during 200 epochs.

The testing is performed on 10,000 image samples from the Microsoft COCO dataset [38] as the cover-image, and 10,000 images of the testing division of the binary CIFAR as the watermark. To demonstrate that the proposed system generalizes the watermarking rules without over-fitting to the training samples, both the testing cover-images and testing watermarks are not used in the training. The peak signal-to-noise ratio (*PSNR*) and bit-error-rate (*BER*) are also respectively used to quantitatively evaluate the fidelity of the marked image and the quality of the watermark extraction in the testing. The *PSNR* is defined as

$$PSNR = 10 log_{10}(\frac{\max(c_i)^2}{MSE(c_i, m_i)}) \quad (8)$$

where *MSE* is mean squared error. The *BER* is computed as the percentage of error bits on the binarization of watermark extraction $w_i^*$. In the testing, the *BER* is zero, indicating that the original and the extracted watermarks are identical. The testing *PSNR* is 39.72 dB, meaning a high fidelity of the marked-images, so that the hidden information cannot be noticed by human vision. A few testing examples with various image content and color are presented in Fig. 10. The residual error showing the absolute difference in each RGB channel between the marked- and the cover-images is also displayed, which demonstrates that the watermark is dispersed over the marked image. This provides extra security to the marked-image. Even when the cover-image is leaked, its subtraction from the marked-image would not reveal the watermark. After the pixel values are rescaled between 0 and 255, the mean of the absolute difference for each RGB channel is computed. The averages over the testing set yield 2.57, 2.10, and 1.63, respectively. The average maxima of the RGB absolute differences are 14.11, 24.79, and 17.08, respectively. These numbers indicate that there are only slightly spiky modification to enable the extraction, but on average the watermark insertion does not alter the channels significantly.

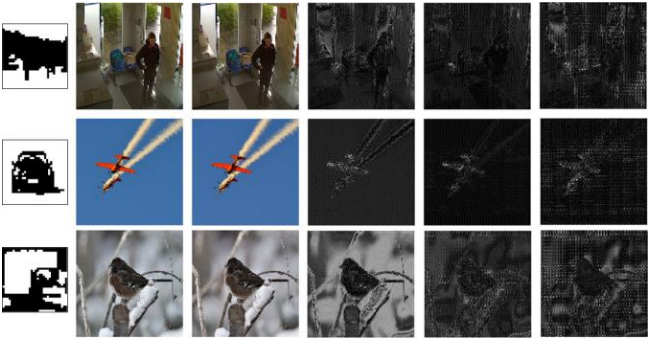

Fig. 10. A few testing examples. First column: the watermark, second column: the cover-image, third column: the marked-image, and fourth, fifth, and sixth columns: the absolute differences of R, G, and B channels between the marked- and the cover-images.

### B. Synthetic Images

To further validate that the watermark embedding and extraction rules are learned without over-fitting, the proposed system is exposed to some extreme cases with synthetic images. In particular, the synthetic situations that are not included during the training process are analyzed, and the results involving blank and random generated images and watermarks are presented.

Fig. 11 shows the results of embedding watermarks into synthetic blank cover-images of red, green, and blue, where the residual errors are increased tenfold. Although the blank cover-images are not included in the training, the proposed system provides promising results. The residual errors display more green color and the blank green marked image displays relatively more noticeable noises than those in other colors, implying that the proposed system modifies the green color slightly more. Applying blank cover-images is known to be extremely difficult in conventional watermarking methods due to the lack of psycho-visual information. However, instead of assigning some unnoticeable portions of visual components as the watermark, the proposed deep learning model learns to apply the correlation between the features of space $W_f$ and the features of the fused space $M$.

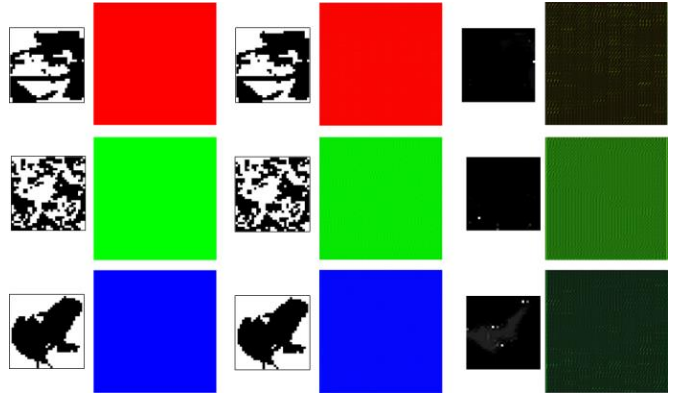

Fig. 11. Embedding watermarks into blank cover-images. First column: the watermark, second column: the blank cover image, third column: the extracted watermark, forth column: the marked-image, and fifth and sixth columns: the residual errors.

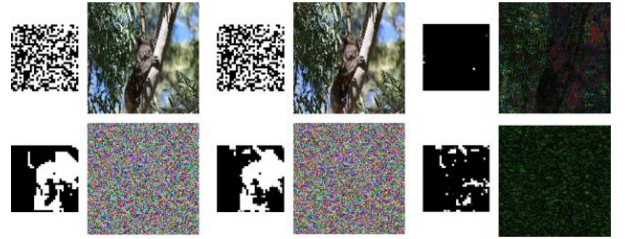

Fig. 12. With noise images. First column: the watermark, second column: the cover-image, third column: the extracted watermark, fourth column: the marked-image, and fifth and sixth columns: the residual errors.

Fig. 12 presents the result of embedding a randomly generated binary image into a natural cover-image, as well as the result of embedding a testing binary watermark into a random color-spotted cover-image. For random watermarks, 10,000 randomly generated bits are tested on 10,000 cover-images from the testing dataset and the average *BER* is 0.36%, which indicates that applying random binary stream as the watermark does not cause problems to the proposed system.

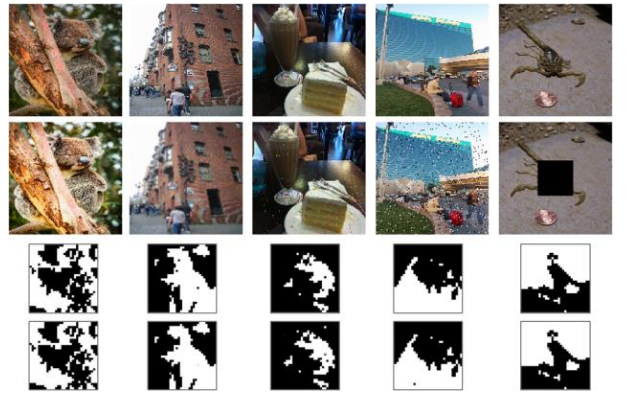

Fig. 13. Visual comparison. First row: marked-images, second row: distorted marked-images, where the distortions from left to right respectively are histogram equalization, Gaussian blur, random noise, salt-and-pepper noise, and cropping, third row: original watermarks, and fourth row: watermark extractions from the distorted marked-images.

When it comes to embedding watermarks into random cover-images, a test of embedding 10,000 watermarks from the testing dataset into 10,000 randomly generated cover-images yields a higher average *BER* of 11.98%. Although the general shape is still recognizable, there are obvious distortions on the watermark extraction. However, in practical applications, embedding a watermark into random noises means that the appearance of the marked-media is noisy and meaningless, so the encryption methods mapping a watermark into random patterns could be used instead.

*C. Robustness*

The robustness of the proposed system against different distortions applied to the marked-image is evaluated by analyzing the distortion tolerance range. Fig. 13 shows some visual comparison between the marked-images and their distortions, as well as between the original watermarks and the watermark extractions from the distorted marked-images.

Quantitatively, distortions with swept-over parameters that control the attack strength are applied on the marked-images produced from the testing dataset. The watermark extraction *BER* caused by each distortion under each parameter is averaged over the testing dataset. Some distortions with swept-over parameters versus the average *BER* are plotted in Fig. 14. Since the proposed system is designed against image-processing attacks and the input to the system is assumed to be pre-processed to rectify the geometric distortions such as rotation, scaling and translation, the responses of the proposed system against some challenging and common image-processing attacks are discussed here.

The extracted watermarks respectively have 10.6%, 7.8%, 32.2%, 11.6%, 46.2%, and 12.3% average *BER* when the distortions are a Gaussian blur with mean 0 and variance 85%, a cropping discarding 65% percent of the marked image, a Gaussian additive noise mean 0 and variance 20%, a JPEG compression with quality factor 10, a 20% random noise, and a 90% salt-and-pepper noise. The proposed system shows high tolerance range on these challenges especially for cropping, salt-and-pepper noise, and JPEG compression. The attacks that randomly fluctuate the pixel values through image channels show higher *BER* including Gaussian additive noise and random modificative noise. However, a 20% Gaussian additive noise or a 20% random modificative noise destroys most of the contents on the marked-image as shown in Fig. 15, and the proposed system responds acceptable performances given a decent distortion parameter, such as 16% *BER* on 10% Gaussian noise.

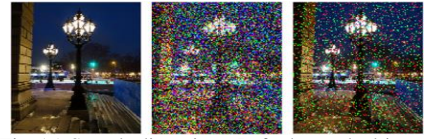

Fig. 15. Sample distortions. Left: the marked-image, middle: after 20% Gaussian additive noise, and right: after 20% random modificative noise.

*D. Comparison*

The proposed system is analytically compared against several state-of-the-art image watermarking methods that incorporate deep neural networks as shown in Table I. Kandi *et al.* [18] proposed to use convolutional neural networks for image watermarking. It applies two deep autoencoders to rearrange a cover-image to a marked-image. To indicate a watermark in the marked-image, the pixels produced by the first auto-encoder represent bit zero and the pixels produced by the second represent bit one. However, the method is a non-blind scheme although achieving robustness. Embedded by increasingly changing an image block to represent a watermark bit, the system in [22] is trained to extract the watermark bits from their corresponding blocks with attack simulation and achieves both blindness and robustness. However, it requires to include the distortions in the training phase for robustness. In reality, we have no way to predict and enumerate all kinds of attacks. To overcome this, our proposed system not only applies deep neural networks to learn the rules of both embedding and extraction, but also intends to achieve blindness and robustness simultaneously without the requirement of the attacks' prior knowledge, and hence has a wider range of applications.

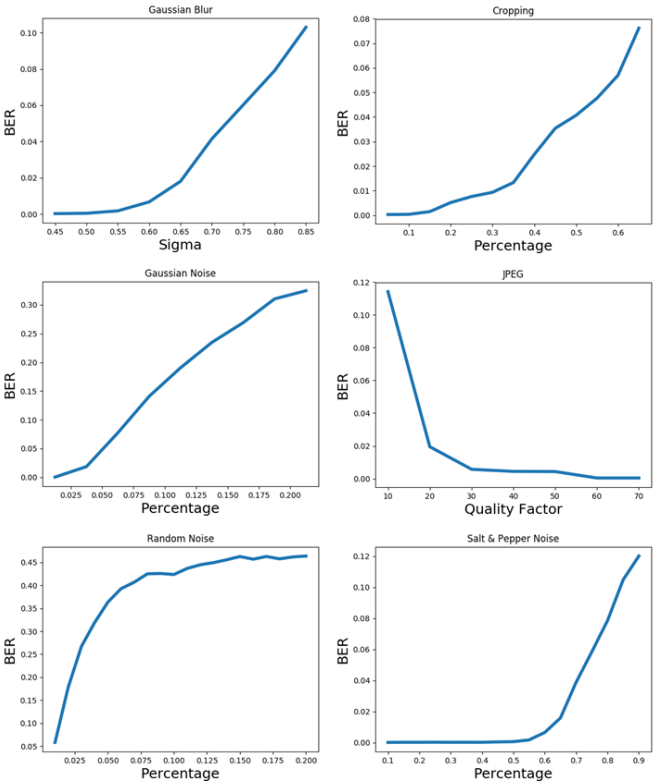

Fig. 14. Distortions with swept-over parameters versus average *BER*.

TABLE I
COMPARISON BETWEEN THE PROPOSED SYSTEM AND STATE-OF-THE-ART IMAGE WATERMARKING METHODS APPLYING DEEP NEURAL NETWORKS

| Method | Function of the deep neural network | Blind | Robust | Concentration |
|---|---|---|---|---|
| [17] | Embedding | no | no | Undetectability |
| [18] | Embedding and extraction | no | yes | Robustness |
| [19] | Embedding and extraction | yes | no | Capacity |
| [20] | Extraction | yes | no | Undetectability |
| [22] | Extraction | yes | yes | Robustness |
| Ours | Embedding and extraction | yes | yes | Robustness |

The proposed system is also quantitatively compared against several related competitors that are blind and robust image watermarking systems. The selection of the competitors considers variation and their concentrations. Mun *et al*. [22] applied convolutional neural networks, and Zong *et al*. [11], Zareian and Tohidypour [39], and Ouyang *et al*. [40] used manually-designed, traditional, and robust methods respectively with different image domains including histogram domain adopting statistical image features, frequency domain, and log-polar domain with summarized image features. All the selected competitors focus on the robustness against image-processing attacks. The testing is performed on the same cover-image sets as well as the same watermarks reported in the references. As the proposed system focuses on common image-processing attacks, the crucial results focusing on this category are presented in Table II, where "/" denotes not applicable, S&P denotes the salt-and-pepper noise, and GF denotes Gaussian filtering.

The proposed system shows advantages by covering more distortions in image-processing attacks and obtaining a lower *BER* under the same distortion parameters. For instance, traditional methods such as manipulating the image histogram cannot tolerate the histogram equalization attack. In addition, the proposed method has a higher tolerance range; for example, [22] and [40] can only extract the watermark with a high JPEG quality of 80 to 90, while the proposed method covers as low as 10. Although the method in [39] focusing on the compression has higher performance on the JPEG, the proposed method outperforms the competitors in all other listed distortions. Remarkably, the competitors tolerate cropping 20% to 30%, while the *BER* is as high as 7.8%% even if 66% of the marked-image is cropped. Finally, under a similar *PSNR*, the proposed method shows its advantages by simultaneously achieving the highest robustness and the highest capacity.

TABLE II
QUANTITATIVE COMPARISON BETWEEN THE PROPOSED SYSTEM AND SOME BLIND AND ROBUST COMPETITORS.

| Method | BER ( % ) under the distortions | | | | | PSNR (dB) | Capacity (bits) |
| --- | --- | --- | --- | --- | --- | --- | --- |
| | HE | JPEG 10 | Cropping 20% | S&P 5% | GF 10% | | |
| [11] | / | 17.50 | 7.06 | 3.51 | 6.33 | 46.63 | 25 |
| [22] | / | / | 6.61 | 7.98 | 4.81 | 38.01 | 1 / block |
| [39] | / | 2.15 | / | 4.94 | 0.21 | 41.00 | 256 |
| [40] | / | / | 7.51 | 9.41 | 27.91 | 36.77 | 24 |
| Ours | 0.43 | 8.16 | 0 | 0.97 | 0 | 39.93 | 1,024 |

## IV. AN APPLICATION: WATERMARK EXTRACTION USING A PHONE CAMERA TO SCAN A SCREEN

To the best of our knowledge, all the methods solving the problem of watermark extraction from camera resample focus on printed papers up to now [23-28]. Applying deep neural networks for watermark extraction from camera resamples of a screen remains unexplored. Although the paper printings sometimes bring noises such as printing quality and bending, the watermark extraction from the resamples of a screen presents a much more challenging task. Besides the noises brought by the camera including geometric distortion, optical tilt, quality degradation, compression, lens distortions, and lighting variation, it introduces much more possible noises from the screen, such as the Moire pattern (i.e., the RGB ripple), the refresh rate of the screen, and the spatial resolution of a monitor (see the examples of camera resamples in Fig. 16). Developing a blind image watermarking system that is simultaneously robust to all of these distortions is extremely difficult. Since our proposed watermarking system is designed to reject all irrelevant noises instead of focusing on certain types of attacks, its application to deal with this problem seems feasible. The outlined process of this application is shown in Fig. 16.

First, an information provider prepares the data by encoding through some error correction coding (ECC) techniques. Then, the marked-image can be obtained by fusing the encoded watermark and the cover-image using the trained embedder network. The marked-image that looks identical to the cover-image is distributed online and displayed on the user's screen. Finally, the user scans the marked-image to extract the hidden watermark by the trained extractor network in our proposed system.

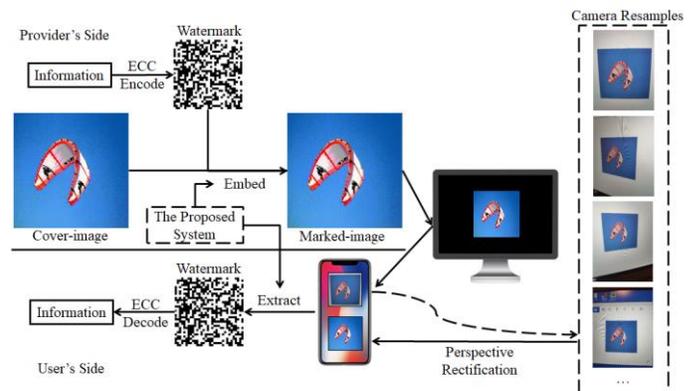

Fig. 16. Process of the application.

The distortions occurred in the application can be divided into two categories: projective and image-processing distortions. The geometric and projective distortions will be rectified by image registration techniques, and the major function of the proposed system in this application is to overcome the pixel-level modifications coming from image-processing distortions, such as the compression, lighting variations, the Moire pattern, and the interpolation errors from the rectification. The autofocus function of a smartphone is utilized.

To simulate a realistic situation, a prototype is developed for a user study, and a 32 × 16 information is used for its clear structure. The user interface (UI) and the sample information are shown in Fig. 17. Reed Solomon (RS) code [41] is adopted as the ECC to protect the information under some *BER*. RS(32,16) is applied to protect each row of the 32 × 16 information, so the encoded information will be a 32 × 32 watermark satisfying the fixed watermarking capacity of the proposed system. In the watermark, each row is a codeword consisting of a data of length 16 and a parity of length 16, and hence can correct up to an error of length 8. Therefore, inside this watermark of length 1,024, up to 256 errors can be corrected if there are no more than 8 errors in each row. Applying half of the bits as the parity, the watermarking payload is 512 bits. As shown in the UI, the prototype only analyzes the region of interest (ROI) in a camera view and hand-taken pictures can hardly be parallel to a screen. Therefore, there exist some geometrical, affine, and perspective distortions, which the proposed system does not concentrate on. Therefore, the image registration technique in [42, 43] is adopted to rectify these distortions before inputting a picture to the proposed system for an extraction. To simplify the prototype as shown in Fig. 18, four corners of the largest contour inside the ROI are used as the reference points. The contoured content is mapped on the bird view plane, and the watermark is extracted from the rectification.

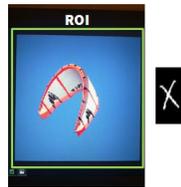

Fig. 17 A prototype and an information. Left: UI of the prototype, and right: the sample information.

Five volunteers were asked to take a few pictures of some marked-images displayed 425px × 425px on a 2,560 × 1,440 screen by the camera of a mobile phone. Two rules were given to the users. First, the entire image should be placed as large as possible inside the ROI. As a prototype for demonstration, this rule facilitates our segmentation that the largest contour inside the ROI is the marked-image, so that this application can focus on the test of the proposed system instead of some complicated segmentation algorithms. In addition, placing the image largely in the ROI helps with the capture of desired details and features for the watermark extraction. Second, the camera should be kept as still as possible. Although the proposed system tolerates some blurring effects, it is not designed to extract watermark in high-speed motion. Fig. 19 presents a few extractions and their corresponding ROIs, where the *BER*s from left to right are 3.71%, 4.98%, 1.07%, 4.30%, and 8.45%, respectively. It can be observed that the closer the picture is taken, the lower the error is. The more parallel between the camera and the screen, the lower the error is. The angle tolerance between the camera and the screen is around 30°. The flashlight brings more errors since it may over- and underexpose some image areas. The flashlight may be turned off in this application since the screen has backlit. There are 20 images in the user's test, and the average *BER* is 5.13%.

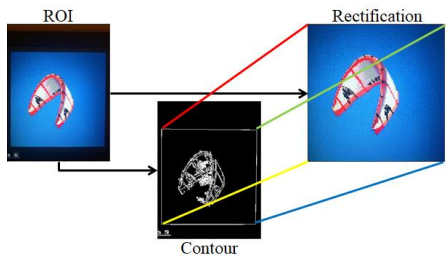

Fig. 18. A marked-image rectification inside a ROI.

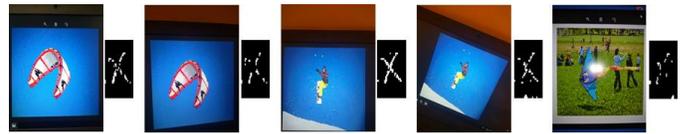

Fig. 19. A few watermark extractions before ECC and the ROIs.

For visual comparison, the displayed sample watermark extractions are the raw result before error correction. After executing RS(32,16), all the watermark extractions in the testing cases can be restored to the original information in Fig. 17 without error. In these tests, the proposed system extracts the watermark within a second as it only applies the trained weights in the extractor network on the marked-image rectification.

## V. Conclusions

This paper introduces an automated image watermarking system using deep convolutional neural networks. The proposed blind image watermarking system achieves its robustness property without requiring prior knowledge of possible distortions on the marked-image. The proposed system constructs an unsupervised deep neural network structure with a novel loss computation for automated image watermarking. Experimental results along with a challenging application of watermark extraction from camera resampled marked-images have confirmed the superiority performance of the proposed system. By exploring the ability of deep neural networks in the task of fusion between the cover-image and the latent spaces of the watermark, the proposed system has successfully developed an image fusion application on image watermarking.

## References

[1] H. Berghel and L. O'Gorman, "Protecting ownership rights through digital watermarking," *Comput.*, vol. 29, no. 7, pp. 101-103, 1996.
[2] R. Caldelli, F. Francesco and R. Becarelli, "Reversible watermarking techniques: An overview and a classification," *EURASIP J. Inform. Security*, vol. 2010, no.1, p. 134546, 2010.
[3] B. Gunjal and R.R. Manthalkar, "An overview of transform domain robust digital image watermarking algorithms," *J. Emerg. Trends in Comput. Inform. Sci.*, vol. 2, no. 1, pp. 37-42, 2010.
[4] I. Cox, M. Miller, J. Bloom, J. Fridrich and T. Kalker, *Digital watermarking and steganography*, Burlington, MA, USA: Morgan Kaufmann, 2008.
[5] F. Y. Shih, *Digital Watermarking and Steganography: Fundamentals and Techniques Second Edition*, Boca Raton, FL, USA: CRC Press, 2017.
[6] X. Kang, J. Huang, Y.Q. Shi, Yan Lin, "A DWT-DFT composite watermarking scheme robust to both affine transform and JPEG compression," *IEEE Trans. Circuits Syst. Video Technol.,* vol.13, no. 8, pp. 776-786, 2003.

[7] A. A. Tamimi, A. M. Abdalla, and O. Al-Allaf, "Hiding an image inside another image using variable-rate steganography," *Int. J. Adv. Comput. Sci. Appl.*, vol. 4, no. 10, pp. 18-21, 2013.

[8] I. J. Cox, J. Kilian, F. T. Leighton, and T. Shamoon, "Secure spread spectrum watermarking for multimedia," *IEEE Trans. Image Process.*, vol. 6, no. 12, pp. 1673-1687, 1997.

[9] F. Y. Shih and X. Zhong, "Intelligent watermarking for high-capacity low-distortion data embedding," *Int. J. Pattern Recognit. AI*, vol. 30, no. 5, p. 1654003 (17 pages), 2016.

[10] T. Pevný, T. Filler, and P. Bas, "Using high-dimensional image models to perform highly undetectable steganography," in *Proc. Int. Workshop Inform. Hiding*, Calgary Canada, Jun. 28 – 30 2010, pp. 161-177.

[11] T. Zong, Y. Xiang, I. Natgunanathan, S. Guo, W. Zhou, and G. Beliakov, "Robust histogram shape-based method for image watermarking," *IEEE Trans. Circuits Syst. Video Technol.*, vol. 25, no. 5, pp. 717-729, 2015.

[12] Y. Qian, J. Dong, W. Wang and T. Tan, "Deep learning for steganalysis via convolutional neural networks," *Media Watermarking, Security, and Forensics*, vol. 9409, p. 94090J, International Society for Optics and Photonics, 2015.

[13] L. Pibre, J. Pasquet, D. Ienco, and M. Chaumont, "Deep learning is a good steganalysis tool when embedding key is reused for different images, even if there is a cover source mismatch," *Electron. Imag.*, vol. 1, no. 11, pp. 1-11, 2016.

[14] H. Sabah and B. Haitham, "Artificial neural network for steganography," *Neural Comput. Appl.*, vol. 26, no.1, pp. 111-116, 2015.

[15] S. B. Alexandre and C. J. David, "Artificial neural networks applied to image steganography," *IEEE Latin Amer. Trans.*, vol. 14, no. 3, pp. 1361-1366, 2016.

[16] J. Robert, V. Eva, and K. Martin, "Neural network approach to image steganography techniques," *Mendel*, pp. 317-327. Springer, 2015.

[17] W. Tang, S. Tan, B. Li, and J. Huang, "Automatic steganographic distortion learning using a generative adversarial network," *IEEE Signal Process. Lett.*, vol. 24, no. 10, pp. 1547-1551, 2017.

[18] H. Kandi, D. Mishra and S. R. S Gorthi, "Exploring the learning capabilities of convolutional neural networks for robust image watermarking," *Comput. & Security*, vol. 65, no. C, pp. 247-268, 2017.

[19] S. Baluja, "Hiding images in plain sight: deep steganography," in *Proc. Adv. Neural Inform. Process. Syst.*, Long Beach CA, 2017, pp. 2069-2079.

[20] D. Li, L. Deng, B. B. Gupta, H. Wang, and C. Choi, "A novel CNN based security guaranteed image watermarking generation scenario for smart city applications," *Inform, Sci.*, vol. 479, no. 4, pp. 432-447, Apr. 2019.

[21] N. Papernot, P. McDaniel, S. Jha, M. Fredrikson, Z.B. Celik, and A. Swami, "The limitations of deep learning in adversarial settings," in *Proc. IEEE Eur. Symp. Security Privacy*, Saarbrucken Germany, Mar. 2016, pp. 372-387.

[22] S. M. Mun, S. H. Nam, H.U. Jang, D. Kim and H. K. Lee, "Finding robust domain from attacks: A learning framework for blind watermarking," *Neurocomputing*, vol 337, pp. 191-202, 2019.

[23] A. Pramila, A. Keskinarkaus and T. Seppänen, "Camera based watermark extraction-problems and examples," in *Proc. Finnish Signal Process. Symp.*, Tampere Finland, 2007.

[24] A. Pramila, A. Keskinarkaus, and T. Seppänen, "Increasing the capturing angle in print-cam robust watermarking," *J. Syst. Softw.*, vol. 135, pp.205-215, 2018.

[25] A. Katayama, T. Nakamura, M. Yamamuro, N. Sonehara, "New high-speed frame detection method: Side trace algorithm (sta) for i-appli on cellular phones to detect watermarks," in *Proc. 3rd Int. Conf. Mobile Ubiquitous Multimedia*, College Park MD, Oct. 2004, pp. 109–116.

[26] W. G. Kim, S. H. Lee and Y. S. Seo, "Image fingerprinting scheme for print-and–capture model." in *Proc. Pacific-Rim Conf. Multimedia*, Heidelberg Berlin Germany, Nov. 2006, pp. 106–113.

[27] T. Yamada and M. Kamitani, "A method for detecting watermarks in print using smart phone: finding no mark." in *Proc. 5th Workshop on Mobile Video*, *ACM*, Oslo Norway, Feb. 2013, pp. 49–54.

[28] L. A. Delgado-Guillen, J. J. Garcia-Hernandez and C. Torres-Huitzil, "Digital watermarking of color images utilizing mobile platforms," in *Proc. 2013 IEEE 56th Int. Midwest Symp. Circuits Syst.*, IEEE, Columbus Ohio, Aug. 2013, pp. 1363-1366.

[29] J. Y. Zhu, T. Park, P. Isola, and A. A. Efros, "Unpaired image-to-image transaction using cycle-consistent adversarial networks," in *Proc. IEEE Int. Conf. Comput. Vis.*, Venice Italy, 2017, pp. 2223-2232.

[30] H. Li and X. J. Wu, "DenseFuse: A fusion approach to infrared and visible images," *IEEE Trans. Image Process.*, vol. 28, no. 5, pp. 2614-2623, 2019.

[31] G. E. Hinton and R. R. Salakhutdinov, "Reducing the dimensionality of data with neural networks," S*ci.*, vol. 313, no. 5786, pp. 504-507, 2006.

[32] S. Rifai, P. Vincent, X. Muller, X. Glorot, and Y. Bengio, "Contractive auto-encoders: Explicit invariance during feature extraction," in *Proc. 28th Int. Conf. Mach. Learn.*, Bellevue Washington, Jun. 2011, pp. 833-840.

[33] Y. LeCun, L. Bottou, G. B. Orr, and K. R. Müller, "Efficient backprop," in *Neural Networks: Tricks of the Trade*, Germany: Springer, 2012, pp. 9-48.

[34] C. Szegedy, S. Ioffe, V. Vanhoucke, and A.A. Alemi, "Inception-v4, inception-resnet and the impact of residual connections on learning," in *Proc. AAAI Conf. AI*, San Francisco CA, Feb. 2017, pp. 4278-4284.

[35] O. Russakovsky, J. Deng, H. Su, J. Krause, S. Satheesh, S. Ma, Z. Huang, A. Karpathy, A. Khosla, M. Bernstein, and A.C. Berg, "Imagenet large scale visual recognition challenge," *Int. J. Comput. Vis.*, vol. 115, no. 3, pp. 211-252, 2015.

[36] A. Krizhevsky and G. Hinton, "Learning multiple layers of features from tiny images," Univ. Toronto, Toronto, Canada, no. 4, vol. 1, pp. 7, 2009.

[37] D. P. Kingma and J. L. Ba, "Adam: a method for stochastic optimization." in *Proc. Int. Conf. Learn. Representations*, San Diego CA, May 2015, pp. 1–13.

[38] T. Y. Lin, M. Maire, S. Belongie, J. Hays, P. Perona, D. Ramanan, P. Dollár, and C. L.Zitnick, "Microsoft coco: Common objects in context," in *Proc. Eur. Conf. Comput. Vis.*, Zurich Switzerland, Sept. 2014, pp. 740-755.

[39] M. Zareian and H. R. Tohidypour, "Robust quantisation index modulation-based approach for image watermarking," *IEEE Trans. Image Process.*, vol. 7, no. 5, pp. 432-441, 2013.

[40] J. Ouyang, G. Coatrieux, B. Chen, and H. Shu, "Color image watermarking based on quaternion Fourier transform and improved uniform log-polar mapping," *Comput. & Elect. Eng.*, vol. 46, pp. 419-432, 2015.

[41] I. S. Reed and G. Solomon, "Polynomial codes over certain finite fields." *J. Soc. Ind. Appl. Math.*, vol. 8, no.2, pp.300-304, 1960.

[42] L. G. Brown, "A survey of image registration techniques." *ACM Comput. surveys*, vol. 24, no.4, pp.325-376, 1992.

[43] S. Zokai, and G, Wolberg, "Image registration using log-polar mappings for recovery of large-scale similarity and projective transformations." *IEEE Trans. Image Process.*, vol. 14, no. 10, pp.1422-1434, 2005.